%% file: main.tex
\documentclass[11pt]{article}

\usepackage[a4paper, total={16cm, 22cm}]{geometry}

\usepackage[backend=biber,style=numeric-comp,sorting=none, url=false]{biblatex}
\usepackage{setspace}
\usepackage{amsmath}
\usepackage{amsfonts}
\newcommand*\diff{\mathop{}\!\mathrm{d}}
\usepackage[most]{tcolorbox}
\usepackage{wrapfig}
\newcommand\up[1]{\textsuperscript{#1}}
\usepackage{siunitx}

\usepackage{soul}
\usepackage{xcolor}
\sethlcolor{yellow}
\usepackage{placeins}

\begin{document}

\include{paper}

\include{supportinginformation}

\end{document}

%% file: paper.tex
\vspace{7mm}
{\huge\textbf{Computational Determination of Optimal Growth Protocols for Metastable Polymorphs}}

\vspace{5mm}
Simon B. Hollweger\up{1}, Anna Werkovits\up{1}, Tadeas Lesovsky\up{1}, and Oliver T. Hofmann\up{1}*
\vspace{3mm}

\begin{quote}
    \up{1}\textit{Institute of Solid State Physics, Graz University of Technology, Graz 8010, Austria}
\end{quote}

\begin{quote}
    *\textit{Corresponding Author, Email: o.hofmann@tugraz.at}
\end{quote}

\section{Abstract}

The reliable growth of a desired target structure remains a central challenge for organic--inorganic interfaces. Specific interface structures can exhibit properties that are superior compared to those of other possible interface structures, but identifying growth conditions that selectively produce a given surface structure is difficult, particularly when the target structure is thermodynamically metastable. Here, we demonstrate how time-dependent temperature and pressure protocols can be optimized  to promote the high-yield formation of a metastable surface polymorph. 

To this end, we combine kinetic Monte Carlo simulations with a parameterized nucleation-and-growth model and apply optimal control theory to predict growth recipes that maximize the yield of the desired target structure. Applying this approach to a prototypical model of an organic molecules adsorbed on a metal surface, we identify experimentally plausible protocols that guide the system through phase space while avoiding kinetic growth regimes in which formation of the thermodynamically stable structure is favored. Compared to a manually optimized three-step protocol, the optimized control trajectory increases the yield of the desired metastable phase from 73 \% to 97 \% for the same total protocol duration. 

\begin{refsection}
\section{Introduction}

The growth of (especially molecular) crystals is a competition between kinetic and thermodynamic effects. While thermodynamics provide the driving force towards the structure with the lowest Gibb’s free energy, kinetics favor processes that proceed via small energy barriers. Depending on the shape of the energy surface and the growth conditions employed, this commonly leads to metastable samples that are kinetically trapped. \cite{packwoodChemicalEntropicControl2017, werkovitsKineticTrappingChargetransfer2023}
In science and technology, such metastable structures are often desired, since they exhibit properties that are superior to those of the thermodynamically stable structure. For example, in pharmaceutical applications they commonly exhibit higher bio-availability \cite{llinasPolymorphControlPresent2008}, or in organic electronics, they show much larger charge carrier mobilities. \cite{zhen2016TailoringCrystalPolymorphs} However, while it is generally clear how the thermodynamically stable structure can be reliably grown (i.e., employing elevated temperatures and slow deposition rates), the experimental protocol to reliably grow a specific kinetically trapped structure is generally a priori unclear. \cite{zhaoActiveLearningGuidedPolymorph2026} The situation is further complicated by the fact that often multiple competing metastable structures exist (lowering the yield of the desired material), and that, qualitatively, the necessary conditions – reduced temperatures – may lead to unacceptably large experimental timescales. 
To overcome these obstacles, a common strategy is to control the assembly kinetics by using time-varying temperature protocols during growth. \cite{bupathyTemperatureProtocolsGuide2022, abubakar2009ImpactDirectNucleation} The challenge is finding these optimal protocols resulting in a high yield within a reasonable time frame. 
Unfortunately, identifying such protocols experimentally can be both time-consuming and costly. A more efficient approach is to model the assembly kinetics of the system theoretically, and then optimize self-assembly protocols within this model. Going beyond  that optimized generic self-assembly protocols for colloids, polymers, capsids, or abstract particle models,\cite{tangOptimalFeedbackControlled2016, trubianoOptimizationNonEquilibriumSelfAssembly2022, whitelamLearningGrowControl2020}, in this work we formulate the selective growth of a metastable organic–inorganic interface polymorph as an  optimal-control problem with clear, physically-grounded experimental constraints in temperature and molecular partial pressure, using a physically interpretable nucleation-and-growth model fitted to kMC simulations.

Our approach applies optimal control theory \cite{kirkOptimalControlTheory2004} to a parameterized effective nucleation-and-growth model in order to maximize the yield of the desired target structure within a prescribed time frame. To illustrate the proposed method, we apply it to a prototypical model for the growth of an organic monolayer.  The obtained optimized temperature and pressure protocol reduces the assembly time of the metastable target structure from roughly one hour down to 30 minutes with a minimal reduction in yield.

\section{Results and Discussion}

\begin{figure}
    \centering
    \includegraphics{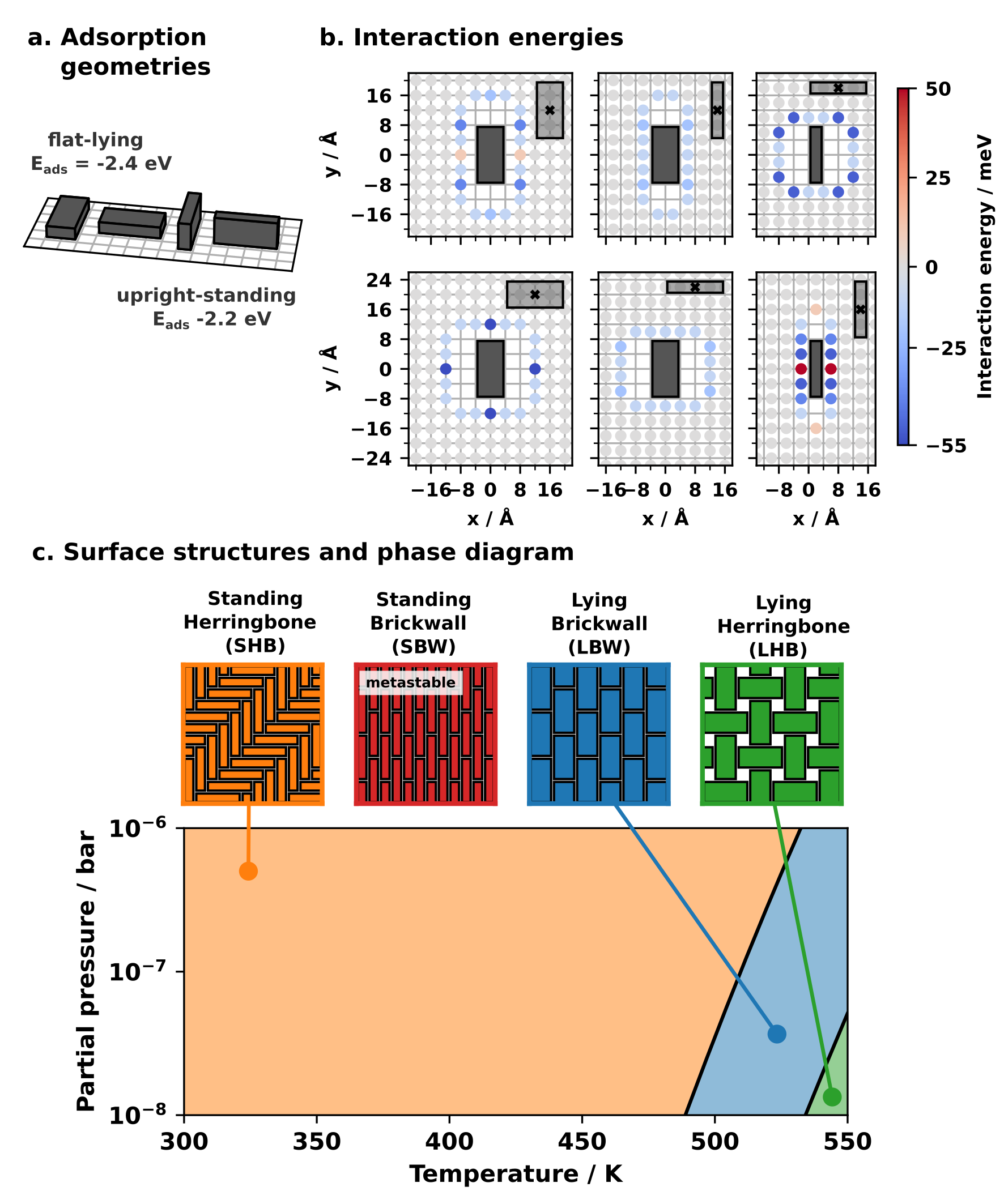}
    \caption{Interface system adapted from Ref. \cite{hollwegerMetastableMonolayerFormation2025a}. a) possible flat lying and upright standing adsorption geometries and energies of the molecules. b) lateral interaction energies of the molecules. c) The four possible surface structures and the corresponding temperature-pressure phase diagram.}
    \label{fig:simon_paper_1}
\end{figure}

To illustrate our method, we consider a prototypical model of $\pi$-conjugated organic molecules on a metal surface. Such systems are known for their extensive polymorphism already in the first layer. \cite{jones2016SubstrateInducedThinFilmPhases} Typical polymorphs include either flat-lying or upright-standing layers \cite{eggerChargeTransferOrganic2020, werkovitsKineticTrappingChargeTransfer2024, roscioniPentaceneCrystalGrowth2018}, packed e.g. in herringbone or brickwall motifs. \cite{puschnig2017EnergyOrderingMolecular, wiessner2012ElectronicGeometricStructure}  These different molecular orientations can give rise to substantially different electronic properties. \cite{willenbockel2013EnergyOffsetsMolecular, coropceanu2007ChargeTransportOrganic, ambrosch-draxlRolePolymorphismOrganic2009} Here, we abstract this class of systems as building blocks that cover either a large area with a large adsorption energy (representing   flat-lying molecules) or a small area with a small adsorption energy (representing upright-standing molecules) (as shown in Figure \ref{fig:simon_paper_1}a), with interactions between the building blocks that are reminiscent of $\pi-\pi$ attraction or electrostatic repulsion, as (Figure \ref{fig:simon_paper_1}b).  This leads to four potential polymorphs, that are shown in Figure \ref{fig:simon_paper_1}c: (i) A phase of lying molecules in herringbone motif (LHB), that will play no further role in our consideration.; (ii) A phase of lying molecules in brickwall motif (LBW), that is thermodynamically stable at high temperatures and low pressures. (iii) A phase of standing molecules in herringbone (SHB) motif, that is stable at low temperatures and high pressures, and (iv) a phase of standing molecules in brickwall (SBW) motif, that is only metastable. For the sake of this work, this metastable structure shall be the target polymorph whose yield will be maximized during growth.

The first step to determine optimized growth protocols is to obtain a reliable model describing the phase transition rates between the different polymorphs of the system of interest, as well as their explicit dependence on the growth conditions. In a typical physical vapor deposition experiment, these conditions would be the temperature $T$ and the deposition rate (given by the partial gas pressure $p$) of the organic molecules. 

We model phase transitions between different phases of a system using a growth-nucleation model with the mathematical form
\begin{equation}
\label{eq:paperI_ode}
    \frac{\diff c_i}{\diff t} = \sum_{j=1}^{N} \Big[ n_{ji}(T, p) c_j + k_{ji}(T, p) l_{ji} \Big]
\end{equation}

Where $c_i$ is the fraction of the surface occupied by polymorph $i$, $n_{ji}$ the nucleation rate of polymorh $i$ in polymorph $j$ (i.e., how frequently nuclei of polymorph $i$ appears in polymorph $j$), $k_{ji}$ the rate constant for the growth of a nucleus of structure $i$ incorporated in structure $j$, and $l_{ji}$ the contact length between polymorph $i$ and $j$. The physical intuition behind the latter term is that the absolute growth of an existing nucleus is proportional to the length of its exposed grain boundary, as these are the only regions where incoming molecules can attach in the appropriate configuration to contribute to grain growth.\cite{paris2018KineticControlMolecular} 
The temperature and pressure dependence of both rate constants $k_{ij}\left(T,p\right)$ and $n_{ij}\left(T,p\right)$ are approximated by Arrhenius-like expressions of the form 
\begin{align}
\label{eq:paperoct_growthrate}
k_{ij}\left(T,p\right)=f_{ij}\exp{\left(-\frac{\mathrm{\Delta}G_{ij}\left(T,p\right)}{k_BT}\right)}-f_{ji}\exp{\left(-\frac{\mathrm{\Delta}G_{ji}\left(T,p\right)}{k_BT}\right)}
\end{align}
\begin{align}
\label{eq:paper_oct_nucleation_rate}
n_{ij}\left(T,p\right)=f_{ij}^{\mathrm{nucl}}\exp{\left(-\frac{\Delta G_{ij}^{\mathrm{nucl}}(T,p)}{k_BT}\right)},
\end{align}
with $\mathrm{\Delta}G_{ij}$ being the free energy barrier for the transition from $i$ to $j$ and $f_{ij}$ the corresponding attempt frequency. Likewise, $f_{ij}^{\mathrm{nucl}}$ is the attempt rate to form a (critical) nucleus, while $\Delta G_{ij}^{\mathrm{nucl}}$ describes the associated free-energy barrier. This barrier accounts for the probability that a newly formed nucleus reaches and exceeds the critical nucleus size, thereby becoming stable against spontaneous dissolution.

To obtain realistic parameters for our model, we took elementary rates (i.e., attempt rates and effective barriers for diffusion, reorientation, adsorption and diffusion of individual molecules) from Ref. \cite{hollwegerMetastableMonolayerFormation2025a} and performed kinetic Monte Carlo (kMC) simulations on a fine grid of different ($T$,$p$)-points, monitoring the emergence and disappearance of the individual phases over time for different growth conditions. All the parameters appearing in Equations \ref{eq:paperI_ode}--\ref{eq:paper_oct_nucleation_rate} were then obtained using a least-squares fitting procedure. Details about the kMC simulations and the fitting procedure are provided in the Supporting Information.

In an earlier work \cite{hollwegerMetastableMonolayerFormation2025a}, we have shown that it is  possible to start with the thermodynamically stable structure at $T=300 \mathrm{~K}$ (the SHB phase) and convert to the target metastable SBW phase by first increasing the temperature (thus converting to LBW). If the system is fully covered with this low-coverage LBW phase and is cooled down to a temperature-pressure region where standing monolayers are thermodynamically preferred, the metastable SBW phase emerges instead of the stable SHB phase because of the structural similarity between the LBW and the metastable SBW phase. However, the process is highly sensitive to the growth conditions, and identifying suitable conditions requires tedious manual optimization. 
The difficulty of growing the metastable SBW phase in high yield is to reach the temperature-pressure regime that kinetically favors the LBW-SBW transition before a significant amount of the stable SHB phase is accumulated. 

To determine how efficiently we can grow our target structure within a given timeframe, we first need to establish the starting and end points of the growth protocol: We want to start with a system that is in thermodynamic equilibrium (at that time) and completely in the LBW phase, because we know that this can be efficiently converted into our target structure. Here, we chose a slightly elevated temperature of $T_i = 525 \mathrm{~K}$ at a pressure of $p_i = {10}^{-7} \mathrm{~bar}$. At the end of the growth protocol, we want to be at room temperature ($T_\mathrm{f} = 300 \mathrm{~K}$), with a residual pressure $p_\mathrm{f} = {10}^{-7} \mathrm{~bar}$. To keep the protocols realistic, we limited the permitted changes in temperature and pressure to $1 \mathrm{~Ks^{-1}}$ and to one order of magnitude in pressure per 100 s (corresponding to $0.01 \log_{10}(\mathrm{bar})\mathrm{s}^{-1}$), respectively. The total time for the protocol is fixed to a time $t_\mathrm{f}=1800~$s. 

\begin{figure}
    \centering
    \includegraphics[width=0.5\textwidth]{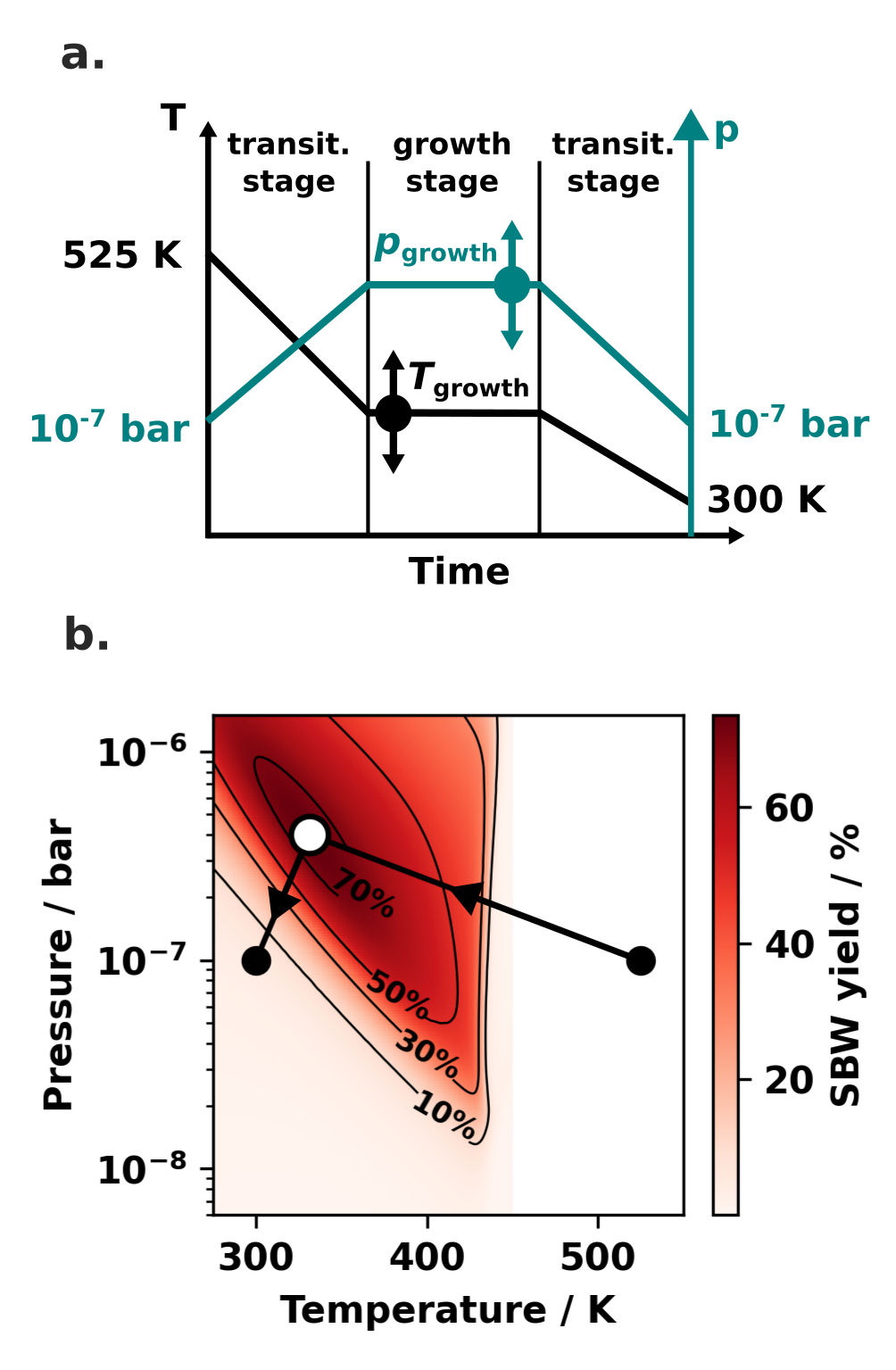}
    \caption{Manually designed 3-stage protocol. a) Schematic depiction of the temperature and pressure protocols. For optimization $T_\mathrm{growth}$ and $p_\mathrm{growth}$ are varied. b) Obtained yield of the metastable SBW structure for a protocol duration of $t_\mathrm{f} = 1800 ~ \mathrm{s}$ for different growth points ($T_\mathrm{growth}$, $p_\mathrm{growth}$). The black line is the best 3-stage protocol found, with a maximum yield of 73 \%}
    \label{fig:fig2}
\end{figure}

To establish a baseline for how well we would be able to determine an efficient growth protocol manually, we start by assuming that such a protocol needs three steps: first, a transition stage changing the growth parameters as fast as possible from the initial conditions to a to-be-determined "growth point" ($T_\mathrm{growth}$, $p_\mathrm{growth}$); second, a period where we keep the conditions constant at ($T_\mathrm{growth}$, $p_\mathrm{growth}$), and third, a return transition stage to the fixed final temperature and pressure of 300 K and ${10}^{-7} \mathrm{~bar}$. The protocol is schematically depicted in Figure \ref{fig:fig2}a. 

Systematically varying the parameters $T_\mathrm{growth}$ and $p_\mathrm{growth}$ for a given $t_\mathrm{f}$ of 1800 s, we determine the resulting yield of the target polymorph in Figure \ref{fig:fig2}b. The best protocol that can be found with this naïve approach provides a yield of 73 \%; the remaining 27 \% of the surface is covered with the (here undesired) thermodynamically stable phase. 

\begin{figure}
    \centering
    \includegraphics[width=11cm]{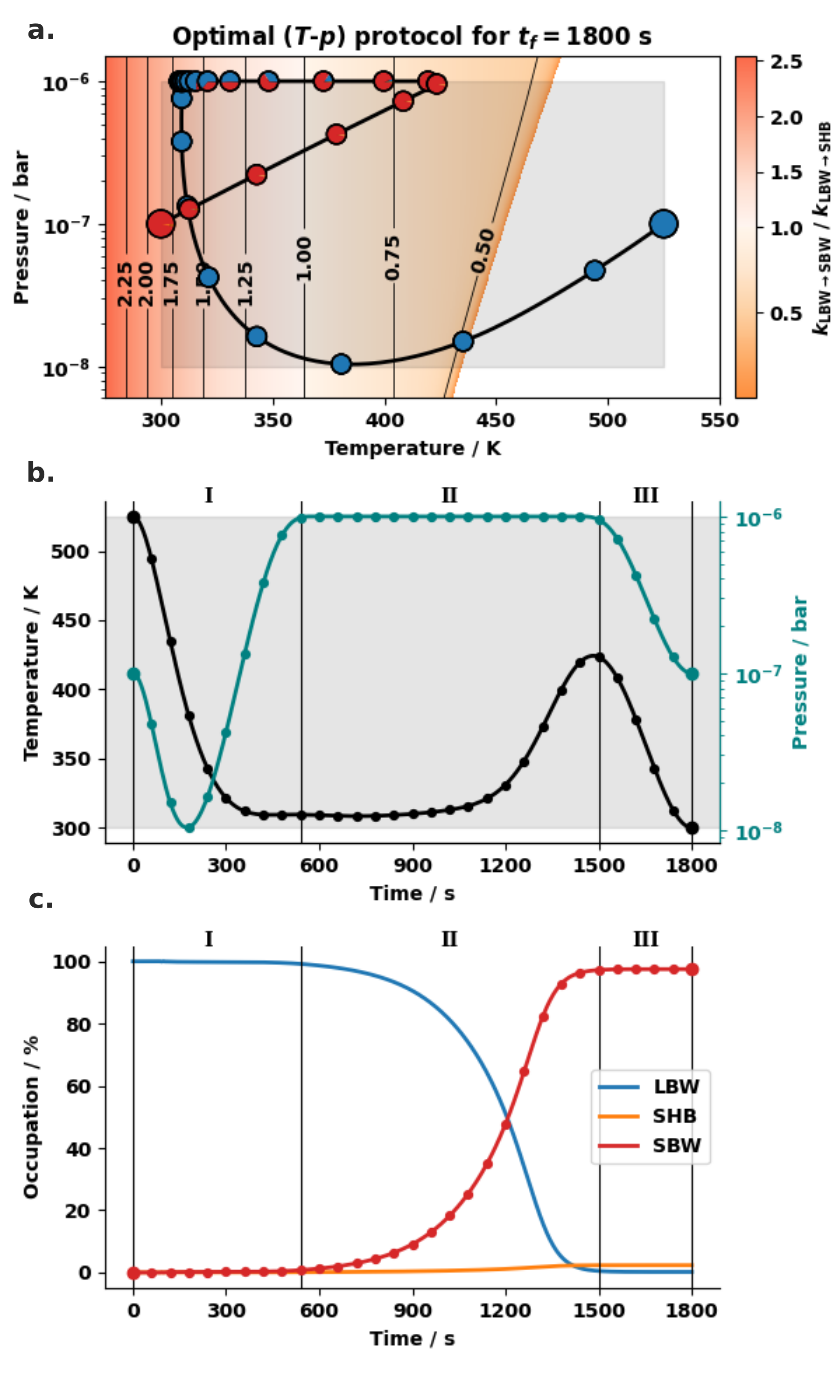}
    \caption{a) Optimal control protocol for a protocol duration of $t_\mathrm{f}=1200$ s (black solid line). The colormap in the background is the ratio $k_\mathrm{LBW\to SBW} / k_\mathrm{LBW \to SHB}$ between the two growth rates of the metastable SBW and the stable SHB structure. The pie charts along the protocol line are snapshots taken every 60 s of the surface occupations at this specific time. b) Optimal temperature and pressure control plotted over time. The markers are placed every 60 s. c) Occupations of the three structures in the system plotted over time. At 1800 s the maximum yield of 97 \% occupation in the SBW structure is reached.}
    \label{fig:fig3}
\end{figure}

To find an improved, optimal growth protocol, we employ a framework based on Optimal Control theory (OCT). In a nutshell, OCT is a mathematical framework that determines the optimal control protocol function to maximize a user-defined objective, in our case, the yield of the metastable SBW surface polymorph. Details on the implementation are provided in the Methods section and the Supporting Information. The boundary condition for this protocol, i.e. the starting and final conditions as well as limits on the maximum changes of pressure and temperature are chosen consistent with the manual optimization above. Additionally, we require the time derivatives of the temperature and pressure curves to vanish at the start and end points. This ensures smooth transitions to constant growth conditions before and after the optimized protocol. Furthermore, for the sake of realism, we limit the allowed temperature range to $300 \mathrm{~K} - 525 \mathrm{~K}$ and the pressure range to ${10}^{-8}$ to ${10}^{-6} \mathrm{~bar}$ at all times (shaded area in Figure \ref{fig:fig3}a and b). 

With these constraints we compute optimized temperature-pressure protocols with OCT for the same duration of $t_\mathrm{f} = 1800 \mathrm{~s}$ that we employed for the manual determination. The result is shown in Figure \ref{fig:fig3}a, which depicts both the optimized values of pressure and temperature over time as well as the composition of the system at this time. Each pie chart along the protocol line represents a snapshot taken every 60\ s. The pie charts show the current surface occupation fractions of the target SBW phase (red), the stable SHB phase (orange) and the initial lying LBW phase (blue) at that specific time. Consequently, pie charts positioned closer together indicate slower changing rates in temperature and pressure.  

In the background of Figure \ref{fig:fig3}a the ratio $k_{\mathrm{LBW} \rightarrow \mathrm{SBW}}/k_{\mathrm{LBW} \rightarrow \mathrm{SHB}}$ of the growth rate constant between the LBW phase and the metastable SBW phase $k_{\mathrm{LBW} \rightarrow \mathrm{SBW}}$, and the connector LBW phase and the stable SHB phase $k_{\mathrm{LBW} \rightarrow \mathrm{SHB}}$, is plotted. Values larger than one indicate that the target structure forms faster than the thermodynamically stable structure. We note that this ratio is only an approximative measure for determining which transition from the connector LBW phase is preferred at this $(T,p)$ point, since the effects of the nucleation rates of the SBW and SHB phase and the different grain boundary lengths are neglected here.

Inspecting the optimal protocol depicted in Figure \ref{fig:fig3}a, it becomes obvious that, conceptually, it consists of the same three parts as the manual three-phase protocols from before: First, it transitions from the initial conditions to a growth point, where it remains for a given time before it finally transitions to the final conditions of the experiment. The details, however, are fundamentally different. 

During the first stage, the optimized protocol does not change both $T$ and $p$ at their maximum allowed rates directly toward the high-pressure, low-temperature regime that favors SBW growth, as was done in the manually optimized protocol. Instead, it follows an initially counterintuitive pathway in which the pressure is first lowered and then increased again. Physically, this allows the system to remain in a situation where the LBW phase is stable as long as possible.
Only once the reduction in temperature leads it into the region where the growth of SBW is faster compared to SHB ($k_{\mathrm{LBW} \rightarrow \mathrm{SBW}}/k_{\mathrm{LBW} \rightarrow \mathrm{SHB}}>1$) the optimized protocol changes pressure and temperature to the “growth point” as quickly as possible.





The reason for that is that the growth of the stable SHB structure is lowest at low pressures. Therefore, entering the SHB-stable phase region at lowest possible pressure reduces accumulation of the undesired SHB structure during this transition stage. 

The second stage of the optimized protocol is the growth stage (path segment II in Figure \ref{fig:fig3}b and c). Here the actual formation of the target phase is achieved. The pressure at this stage is kept constant at the highest possible value of ${10}^{-6} \mathrm{~bar}$, where it remains for a relatively long time. During this time, the initial LBW phase is predominately converted into the metastable target phase, with (relatively) little losses to the thermodynamically stable phase. Towards the end of the growth phase, the temperature is gradually increased. Although this increase means that the ratio of conversion into the target structure versus the stable structure deteriorates, it ensures that the conversion out of the initial phase occurs completely before the timeframe of the protocol (1800~s) expires. 
Finally, the last stage is the return stage to the fixed end conditions of 300 K and ${10}^{-7} \mathrm{~bar}$ which is again performed with maximum temperature and pressure changing rates (path segment III in Figure \ref{fig:fig3}b and c). Notably, despite the same boundary conditions, the optimized protocol leads to a yield of ca. 97~\%, improving the 73~\% we achieved manually.

To validate this prediction, which relies on the fitted parameters for the nucleation/growth model, we applied the optimized protocol in the kMC simulation. 

\begin{figure}
    \centering
    \includegraphics[width=11cm]{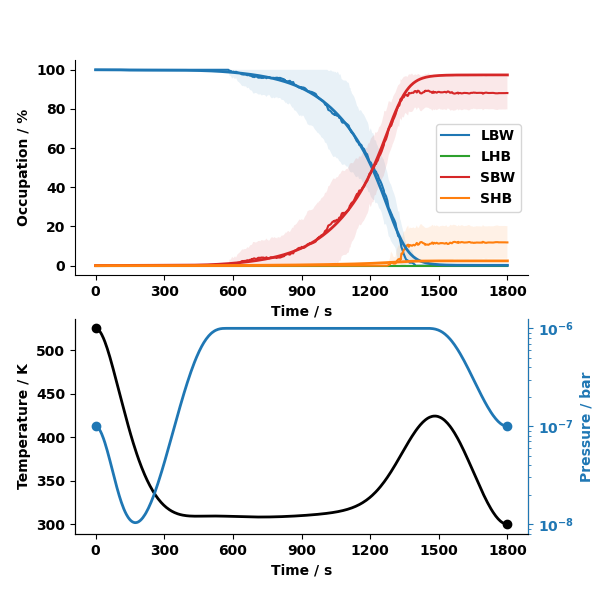}
    \caption{Validation of the obtained optimal protocol for $t_\mathrm{f}=1800 ~ \mathrm{s}$. The upper panel shows the kMC occupation trajectories averaged over 20 independent simulation runs. The shaded area is the standard deviation of the occupations. The smooth curves are the trajectories from the effective growth model. The lower panel shows the applied optimal temperature-pressure protocol.}
    \label{fig:fig4}
\end{figure}

As shown in Figure \ref{fig:fig4}, the predicted yield of the metastable SBW structure from the optimal $T-p$ trajectory agrees very well with the kMC simulation over most of the protocol. Toward the final stage, however, a noticeable deviation emerges between the effective growth model and the kMC simulation. This indicates that the parameterized model does not fully capture the dynamics in this part of the protocol. A possible reason for this discrepancy is that the effective model was parameterized using simulations that did not include configurations with an almost fully SBW-covered surface at temperatures above 400 K. The final part of the optimized protocol therefore lies in an extrapolative regime of the growth model. In this regime the parametrized model appears to underestimate the formation of SHB nuclei, which leads to a reduced yield of the target SBW structure in the kMC simulation.

Nevertheless, the optimized growth protocol obtained from the effective model still correctly identifies conditions under which a high yield of the target SBW structure can be achieved.

\section{Conclusion}

In this work, we have demonstrated that optimized time-dependent growth protocols can be used to selectively steer the formation of a metastable organic surface polymorph. Using a parameterized nucleation-and-growth model fitted to kinetic Monte Carlo simulations, we formulated the growth of competing monolayer structures as an Optimal Control problem in temperature and pressure. This allowed us to identify experimentally tractable protocols that maximize the yield of the desired metastable SBW phase of our model system under realistic bounds on heating, cooling, and pressure variation.

The optimized protocol shows that high-yield growth is not achieved simply by moving as quickly as possible to a single favorable growth point. Instead, the optimized control trajectory exploits the different kinetic growth regimes of the system and guides the growth conditions around regions in phase space where formation of the undesired thermodynamically stable SHB structure is favored. In this way, the protocol suppresses losses into the thermodynamic minimum while maintaining efficient conversion toward the metastable target phase.  
This strategy increases the yield of the metastable target structure from 72 \% for a manually optimized three-step protocol to about 97 \% in the effective model system for the same total protocol duration of $t_\mathrm{f}=1800~\mathrm{s}$.

Beyond the specific model system studied here, these results show that metastable structure formation in organic monolayers can be rationally controlled by combining effective kinetic growth models with Optimal Control Theory. This is particularly relevant for organic electronics, where the microscopic structure of the organic--inorganic interface plays an important role in determining device performance. The ability to selectively grow desired interface polymorphs therefore provides a route toward more controlled interface design and may help overcome current limitations in the fabrication of high-performance organic electronic devices.

\section{Computational Methods}

\subsection{Kinetic Monte Carlo simulations}

In kinetic Monte Carlo \cite{andersenPracticalGuideSurface2019, bortz1975NewAlgorithmMonte, gillespieGeneralMethodNumerically1976}, the complex molecular dynamics of the interface is coarse-grained into a stochastic sequence of elementary processes, such as diffusion, reorientation, adsorption and desorption. Since the present simulations are performed under time-dependent temperature and pressure conditions, the residence time $\Delta t$ is drawn from an inhomogeneous Poisson distribution \cite{jansen1995MonteCarloSimulations} and is determined by solving
\begin{equation}
    \ln(1- r) + \int_0^{\Delta t} k_\mathrm{tot}(t + \Delta t') \diff \Delta t' = 0,
\end{equation}
where $r \in (0, 1)$ is a uniformly distributed random number and $k_\mathrm{tot}(t)$ si the time-dependent total rate of all available processes. For details, see Ref. \cite{hollwegerMetastableMonolayerFormation2025a}. After the time increment has been determined, the executed process is selected randomly according to the relative elementary rates at the updated simulation time.

The adsorption rate is given by the impingement rate of gas-phase molecules onto a kMC lattice cell, \cite{reuterFirstprinciplesKineticMonte2006}
\begin{equation}
    k_\mathrm{ads} = \frac{pA_\mathrm{uc}}{\sqrt{2 \pi m k_\mathrm{B}T}}
\end{equation}
where $p$ is the partial pressure of the molecular gas, $A_\mathrm{uc}$ is the area of one kMC lattice cell, $m$ is the molecular mass, $T$ is the temperature, and $k_\mathrm{B}$ the Boltzmann constant. 

The corresponding desorption rate is approximated by imposing detailed balance on the adsorption-desorption reaction channel. The ratio between adsorption and desorption is chosen such that the equilibrium occupation follows the corresponding Boltzmann factor. With the adsorbed-state energy approximated by the sum of the isolated adsorption energy and the lateral interaction energy, the desorption rate can be written as \cite{reuterFirstprinciplesKineticMonte2006}
\begin{equation}
    k_\mathrm{des} = k_\mathrm{ads} \exp \left( \frac{E_\mathrm{ads} + E_\mathrm{int} - \mu(T, p)}{k_\mathrm{B} T} \right)
\end{equation}
where $E_\mathrm{ads}<0$ is the adsorption energy of an isolated molecule, $E_\mathrm{int}$ is the lateral interaction energy with neighboring molecules, an $\mu(T, p)$ is the chemical potential of the molecular gas. In this work, $\mu(T,p)$ is approximated by the ideal-gas chemical potential. 

The rates of on-surface processes, such as diffusion and reorientation, are described by Arrhenius-type expressions,

\begin{equation}
    k_\mathrm{on\text{-}surf}{} = f \exp{\left(-\frac{\Delta E + \Delta\Delta E_\mathrm{int}}{k_\mathrm{B}T} \right)}
\end{equation}
where $f$ is the attempt frequency, $\Delta E$ is the reference  activation energy barrier, and $\Delta \Delta E_\mathrm{int}$ is a correction term that accounts for changes in lateral interactions between the initial and final states. This correction is determined using the Brønsted-Evans-Polanyi principle.

All kinetic Monte Carlo simulations were performed using the kMC simulation framework \textit{kmos3} \cite{Kmos3org, hoffmannKmosLatticeKinetic2014}. 
The time-scale disparity problem of kMC between fast local diffusion processes and rare growth-relevant events was addressed using the temporal acceleration algorithm proposed by Dybeck et al. \cite{dybeckGeneralizedTemporalAcceleration2017}, as implemented in \textit{kmos3} \cite{Kmos3org, andersenAssessmentMeanfieldMicrokinetic2017}.
The algorithm accelerates the simulation by identifying fast, quasi-equilibrated reaction channels and scaling their rates, thereby increasing the probability of sampling slower, non-equilibrated processes without permanently modifying the physical rate constants.

\subsection{Effective Growth model}

The effective growth model introduced in Equation \ref{eq:paperI_ode} is given by

\begin{equation}
    \frac{\diff c_i}{\diff t} = \sum_{j=1}^{N} \Big[ n_{ji}(T, p) c_j + k_{ji}(T, p) l_{ji} \Big]
\end{equation}
It consists of a nucleation term and growth term. The nucleation term describes the formation of surface structure $i$ from structure $j$, while the growth term accounts for the expansion of already existing domains. The latter is assumed to be proportional to the exposed interfacial length $l_{ij}$ between two surface structures $i$ and $j$. In this work, this interfacial length is approximated as
\begin{equation}
    l_{ij} = c_i^{\alpha_i} c_j^{\alpha_j}.
\end{equation}
which relates the available boundary length $l_{ij}$ to the surface occupation of the two structures $c_i$ and $c_j$ via the phenomenological exponents $\alpha_i$ and $\alpha_j$. Details on the effective growth model and its parameterization to kMC data is given in the Supporting Information.

The temperature and pressure dependence of the effective rate constants $n_{ij}(T, p)$ and $k_{ij}(T, p)$, is described by Arrhenius like expressions. The corresponding functional forms and parameter values are given in the Supporting Information.

\subsection{Optimal Control}

To obtain optimized temperature and pressure trajectories, Optimal Control theory was employed \cite{kirkOptimalControlTheory2004}. The objective was to maximize the final yield of the metastable SBW structure while satisfying the dynamical constraints imposed by the effective growth model. Since the pressure spans several orders of magnitude, the optimization was formulated in terms of the logarithmic pressure variable, $q(t) = \log_{10}p(t)$.
The objective functional is defined as
\begin{equation}
\label{eq:oct_objective}
    J[T, q] = -c_\mathrm{SBW}(t_\mathrm{f}) + \int_0^{t_\mathrm{f}} \left[ \alpha_T \ddot{T}^2(t) + \alpha_q \ddot{q} ^2 (t) \right] \diff t.
\end{equation}
The first term accounts for the target of maximizing the SBW occupation at the final time $t_\mathrm{f}$. Since the optimization is formulated as a minimization problem, this contribution enters with a negative sign. The integral term regularizes the control trajectories by penalizing large curvatures in $T(t)$ and $q(t)$. This promotes smooth temperature and pressure protocols and suppresses abrupt changes. Additionally it allows for imposing initial and final constraints for vanishing first derivatives of $T$ and $q$ at the initial and final time.
The regularization parameters $\alpha_T$ and $\alpha_q$ determine the relative weight of smoothness in the temperature and pressure trajectories, respectively. The optimal control problem is then given by  

\begin{equation}
    \min_{T(t), p(t)} J[T, p]
\end{equation}
subject to the effective growth model 

\begin{equation}
    \frac{\diff c_i}{\diff t} = \sum_{j=1}^{N} \Big[ n_{ji}(T, p) c_j + k_{ji}(T, p) l_{ji} \Big] \quad \text{with} \quad \vec{c}(0) = \vec{c}_0
\end{equation}
with $p(t) = 10^{q(t)}$.
We additionally enforce vanishing initial and final time derivatives of the temperature and pressure to ensure smooth transitions to constant conditions before and after the optimized protocol: 
\begin{align}
    \dot{T}(0) = \dot{T}(t_\mathrm{f}) &= 0\\
    \dot{q}(0) = \dot{q}(t_\mathrm{f}) &= 0.
\end{align}

The controls and their derivatives are constrained by lower and upper bounds. Details on the used values for the constraints are provided in the Supporting Information. 

The optimal control problem was solved using a direct pseudospectral approach. In this method, the continuous optimal control problem is discretized on a collocation mesh, where the state and control trajectories are represented by Lagrange interpolating polynomials. The differential equations, the boundaries, and the constraints are enforced at the collocation points, thereby transforming the original optimal control problem into a finite-dimensional nonlinear programming problem. This nonlinear optimization problem was solved using the Python framework \textit{yapss}. \cite{yapss.readthedocs.io} This framework performs the pseudospectral transcription of the optimal control problem and uses the \textit{Ipopt} \cite{wachter2006ImplementationInteriorpointFilter} solver to solve the resulting finite-dimensional nonlinear optimization problem.
Details of the specific implementation are provided in the Supporting Information.

\section{Acknowledgments}

This research was funded in whole, or in part, by the Austrian Science Fund (FWF) [10.55776/Y1157 and 10.55776/I5170]. For the purpose of open access, the author has applied a CC-BY public copyright license to any Author Accepted Manuscript version arising from this submission. Computational results have been achieved in part using the Vienna Scientific Cluster (VSC). The authors used ChatGPT-5.5 solely to improve the clarity, grammar, and readability of the manuscript. We acknowledge fruitful discussions with S. Matera, M. Deimel, P. Schlosser, B. Ramsauer, C. Wachter, R.K. Berger, L. Hörmann, and J.J. Cartus.

\section{References}

\printbibliography[heading=none]
\end{refsection}

%% file: supportinginformation.tex
\setcounter{section}{0}
\renewcommand{\thesection}{S\arabic{section}}
\setcounter{equation}{0}
\renewcommand{\theequation}{S\arabic{equation}}
\section*{Supporting Information for "Computational Determination of Optimal Growth Protocols for Metastable Polymorphs"}

\vspace{5mm}
Simon B. Hollweger\up{1}, Anna Werkovits\up{1}, Tadeas Lesovsky\up{1}, and Oliver T. Hofmann\up{1}*
\vspace{3mm}

\begin{quote}
    \up{1}\textit{Institute of Solid State Physics, Graz University of Technology, Graz 8010, Austria}

    *\textit{Corresponding Author, Email: o.hofmann@tugraz.at}
\end{quote}

\begin{refsection}
\section{Growth-and-Nucleation model - Detailed description}

The growth model introduced in Equation \ref{eq:paperI_ode} of the main text is given by

\begin{equation}
\label{eq:si_gn_model_ode}
    \frac{\diff c_i}{\diff t} = \sum_{j=1}^N \big(n_{ji} (T, p) c_j +  k_{ji} (T,p) c_i^{\alpha_i} c_j^{\alpha_j}  \big),
\end{equation}

where $c_i$ denotes the surface occupation of polymorph $i$. The quantities $k_{ij}(T, p)$ and $n_{ij}(T, p)$ are temperature- and pressure-dependent effective rate constants describing growth and nucleation contributions associated with the transformation of polymorph $i$ into polymorph $j$, respectively.

The factor $c_i^{\alpha_i} c_j^{\alpha_j}$ is introduced as a phenomenological approximation for the exposed interfacial length between domains of polymorphs $i$ and $j$. This approximation is motivated by the assumption that the growth of an existing domain is approximately proportional to the length of its interface with the surrounding phase. 
An intuitive argument can be obtained by considering the early stages of growth, where domains of the emerging polymorph $j$ are assumed to grow approximately as isolated circular islands within the mother phase $i$. In this limit, the interfacial length is proportional to the perimeter of the islands. Since the occupied area of phase $j$ is proportional to $c_j$, the boundary length scales as $b_j \propto c_j^\frac12$. To account for deviations from ideal circular growth, the exponent $\frac12$ is generalized to a phenomenological exponent $\alpha_j$. Thus, in the early-growth limit, where $c_i \approx 1$, the interface contribution is approximated by $c_j^{\alpha_j}$.  At later stages of growth, domains of the emerging polymorph may coalesce. The morphology is then effectively inverted: small residual domains of the original phase $i$ remain embedded in the surrounding phase $j$. In this limit, the relevant boundary length is governed by the remaining domains of phase $i$ and can be approximated by 
$c_i^{\alpha_i}$. Combining these two limiting cases leads to the phenomenological approximation $l_{ij} \propto c_i^{\alpha_i} c_j^{\alpha_j}$, where $l_{ij}$ denotes the effective interfacial length between polymorphs $i$ and $j$. This expression satisfies two important constraints: it vanishes if either polymorph is absent from the surface, and it is symmetric with respect to exchanging the indices $i$ and $j$, as required for an interfacial length shared by the two phases.

The growth and nucleation rate constants must satisfy certain constraints in order to conserve the total surface occupation. Since the variables $c_i$ denote surface fractions, their sum must remain equal to unity at all times,  
\begin{equation}
    \sum_{i=1}^N c_i(t) = 1
\end{equation}
For this conservation condition to hold, any increase in the occupation of one polymorph must be balanced by an equal decrease in the occupation of another polymorph. 

For the growth term in Equation \ref{eq:si_gn_model_ode} this balancing can be enforced by requiring the growth-rate matrix to be skew-symmetric, 
\begin{equation}
    k_{ij}(T, p) = - k_{ji}(T, p).
\end{equation}
Consequently, transitions from a polymorph to itself do not contribute, and the diagonal elements vanish,
\begin{equation}
    k_{ii}(T, p) = 0.
\end{equation}
For the nucleation term in Equation \ref{eq:si_gn_model_ode}, conservation of the total surface occupation is enforced by collecting the corresponding loss terms on the diagonal of the nucleation-rate matrix. For $i\neq j$, the off-diagonal element $n_{ij}(T, p)$ describes the formation of nuclei of polymorph $j$ at the expense of polymorph $i$. The diagonal element therefore represents the total loss rate out of polymorph $i$, and is defined as 

\begin{equation}
    n_{ii}\left(T,p\right)=-\sum_{j\neq i}{n_{ij}\left(T,p\right)}.
\end{equation}
With this definition, each row of the nucleation-rate matrix sums to zero,
\begin{equation}
    \sum_{j=1}^N n_{ij}(T, p) = 0
\end{equation}
Together with the skew-symmetry of the growth-rate matrix, this ensures conservation of the total surface occupation,
\begin{equation}
    \sum_i \frac{\diff c_i}{\diff t} = 0.
\end{equation}

\section{Growth-and-Nucleation model - Temperature and pressure dependence}

The temperature and pressure dependence of the effective rate constant expressions in Equation \ref{eq:si_gn_model_ode} are approximated by Arrhenius like expressions of the form
\begin{equation}
\label{eq:arrhenius_effective_rate}
    \kappa(T, p; f, \Delta G) = f\exp{\left(-\frac{\Delta{G}(T,p)}{k_\mathrm{B} T}\right)}
\end{equation}
where $f$ is a constant prefactor, $k_\mathrm{B}$ is the Boltzmann constant, and $\Delta G(T, p)$ is the Gibbs free energy barrier. The latter depends on temperature $T$ and partial pressure $p$ and is approximated through the chemical potential of the gas reservoir. In the present model, the barrier is written as
\begin{equation}
    \Delta G(T, p) = \Delta E - \mu_\mathrm{gas}(T, p) \Delta \theta
\end{equation}
where $\Delta E$ is the energetic contribution to the activation energy barrier, $\mu_\mathrm{gas}(T, p)$ is the chemical potential of the gas-phase molecule, and $\Delta \theta$ specifies the number of gas-phase molecules involved in the elementary structural change. Thus, $\Delta \theta$ accounts for whether the process exchanges molecules with the gas reservoir. For processes that do not involve gas-phase molecules, $\Delta \theta = 0$, and the barrier reduces to the purely energetic contribution,
\begin{equation}
    \Delta G(T, p) = \Delta E.
\end{equation}
In this work the values of  $\Delta \theta$ are restricted to $\Delta \theta \in \{-1, 0, 1\}$ for simplicity.  Thus, each process is treated either as independent of the gas reservoir, $\Delta \theta =0$, or as involving the incorporation of one gas-phase molecule $\Delta \theta = \pm1$. This choice corresponds to treating $\Delta \theta$ as an effective stoichiometric coefficient for the exchange of molecules with the gas reservoir. The fixed integer values provide an accurate description of the kinetic Monte Carlo data within the considered model. The values used for the different processes are given in Expression \ref{eq:theta_values}.

The gas-phase chemical potential $\mu_\mathrm{gas}(T, p)$ is described using an ideal-gas approximation, which is commonly employed for molecule-surface interface systems. \cite{reuterCompositionStructureStability2001, rogalInitioAtomisticThermodynamics2007a} Following our previous work \cite{hollwegerMetastableMonolayerFormation2025a}, only translational and rotational contributions to the molecular partition function are included, while vibrational contributions are neglected, as similarly done in literature. \cite{wachterPhaseDiagramsOrganic2025} 

For a nonlinear rigid molecule, this gives

\begin{equation}
\begin{aligned}
     \mu_{\mathrm{gas}}(T, p) =& \mu_\mathrm{trans}(T, p) + \mu_\mathrm{rot}(T) = \\ =&-k_\mathrm{B} T \ln \left[ \left( \frac{2 \pi m}{h^2} \right)^\frac32 \frac{(k_\mathrm{B} T)^\frac52}{p} \frac{\sqrt{\pi I_1 I_2 I_3}}{\sigma} \left(  \frac{8 \pi^2 k_\mathrm{B}T}{h^2} \right)^\frac32 \right].
\end{aligned}
\end{equation}

Here, $m$ is the molecular mass, $h$ is Planck's constant, $I_1$, $I_2$ and $I_3$ are the principal moments of inertia, and $\sigma$ is the rotational symmetry number \cite{hollwegerMetastableMonolayerFormation2025a, cramerEssentialsComputationalChemistry2013, mcquarrieStatisticalMechanics2000}. This expression can be written in the compact form
\begin{equation}
    \mu_\mathrm{gas}(T, p) = - k_\mathrm{B}T \ln \left[ C_\mathrm{gas} ~(k_\mathrm{B}T)^4 p^{-1} \right]
\end{equation}
where the molecule-specific constant is
\begin{equation}
    C_\mathrm{gas} = \left( \frac{2 \pi m}{h^2} \right)^\frac32  \frac{\sqrt{\pi I_1 I_2 I_3}}{\sigma} \left(  \frac{8 \pi^2 }{h^2} \right)^\frac32.
\end{equation}

Equivalently to our previous work \cite{hollwegerMetastableMonolayerFormation2025a} we use 9-10 anthraquinone as a reference molecule. The used physical quantities are listed in Table \ref{tab:anthra_params}.

\begin{table}[ht]
\centering
\caption{Physical parameters of 9-10 anthraquinone for chemical potential calculation}
\label{tab:anthra_params}
\begin{tabular}{|l|l|l|}
\hline
Parameter & Symbol & Value \\
\hline
Mass & $m$ & $3.458 \times 10^{-25}\,\mathrm{kg}$ \\

Moment of inertia & $I_1$ & $7.636 \times 10^{-45}\,\mathrm{kg\,m^2}$ \\
                  & $I_2$ & $1.895 \times 10^{-44}\,\mathrm{kg\,m^2}$ \\
                  & $I_3$ & $2.659 \times 10^{-44}\,\mathrm{kg\,m^2}$ \\

Symmetry number & $\sigma$ & 4 \\
\hline
\end{tabular}
\end{table}

Inserting this expression into Equation \ref{eq:arrhenius_effective_rate} yields
\begin{equation}
    \kappa(T, p; f, \Delta E, \Delta \theta) = \bar{f}~ T^{-4 \Delta \theta} p^{\Delta \theta}   \exp{\left(-\frac{\Delta E  }{k_\mathrm{B} T}\right)}
\end{equation}
with $\bar{f} = f  ~(C_\mathrm{gas} k_\mathrm{B}^4)^{-\Delta \theta}$. 

In the present work, this expression was simplified further by absorbing the comparatively weak algebraic temperature dependence $T^{-4 \Delta \theta}$ into the fitted Arrhenius parameters. This yields the effective rate expression
\begin{equation}
    \kappa(T, p; f, \Delta E, \Delta \theta) = \bar{f}~ p^{\Delta \theta}   \exp{\left(-\frac{\Delta E  }{k_\mathrm{B} T}\right)}
\end{equation}
where $\bar{f}$ and $\Delta E$ should be interpreted as effective fit parameters. 

With this expression, the effective growth-rate constant is given as the difference between two opposing contributions,
\begin{equation}
\label{eq:growth_rate_const}
    k_{ij}(T, p) = \kappa(T, p; f^\mathrm{grow}_{ij}, \Delta E^\mathrm{grow}_{ij}, \Delta \theta^\mathrm{grow}_{ij}) -  \kappa(T, p; f^\mathrm{grow}_{ji}, \Delta E^\mathrm{grow}_{ji}, \Delta \theta^\mathrm{grow}_{ji})
\end{equation}
where the parameters are collected in the corresponding matrices $f_{ij}$ and $\Delta E_{ij}$ and $\Delta \theta_{ij}$. 
The nucleation contribution is described by a single Arrhenius-like term, 
\begin{equation}
    n_{ij}(T, p) = \kappa(T, p; f^\mathrm{nucl}_{ij}, \Delta E^\mathrm{nucl}_{ij}, \Delta \theta^\mathrm{nucl}_{ij}).
\end{equation}

\section{Least-squares parametrization}

The effective growth model is parametrized using kinetic Monte Carlo simulation data of the interface system introduced in the main text. The system contains four distinct surface polymorphs, with the lying brickwall structure (LBW) acting as a connector phase that enables the growth of the metastable standing brickwall structure (SBW). However, kMC simulations initialized with a fully occupied LBW surface did not show any significant yield of the lying herringbone structure; see Figures \ref{fig:fig250_fit} - \ref{fig:fig525_fit}. Therefore, the lying herringbone structure is omitted from the effective model for simplicity.

The resulting reduced model therefore consists of three states: the connector structure LBW, the metastable target structure SBW, and the thermodynamically stable standing structure SHB. The corresponding state vector is defined as

\begin{equation}
    \vec{c} = \left(\begin{array}{cc}
         c_\mathrm{LBW}  \\
         c_\mathrm{SBW}  \\
         c_\mathrm{SHB}  \\
    \end{array} \right) =  \frac{1}{A_\mathrm{tot}}\left(\begin{array}{cc}
         A_\mathrm{LBW}  \\
         A_\mathrm{SBW}  \\
         A_\mathrm{SHB}  \\
    \end{array} \right)
\end{equation}
where $c_i$ denotes the surface occupation of polymorph $i$, $A_i$ is the surface area occupied by this polymorph, and $A_\mathrm{tot}$ is the total surface area.
The model contains $3 \times3$ parameter matrices for the nucleation contribution $f^\mathrm{nucl}_{ij}, \Delta E^\mathrm{nucl}_{ij}, \Delta \theta^\mathrm{nucl}_{ij}$ , and for the growth contribution, $f^\mathrm{grow}_{ij}, \Delta E^\mathrm{grow}_{ij}, \Delta \theta^\mathrm{grow}_{ij}$.
As discussed above, the effective stoichiometric coefficients entering the chemical-potential contribution are restricted to $\Delta \theta_{ij} \in \{-1, 0, 1\}$. For transitions from the connector phase LBW to one of the standing structures, SBW or SHB, the coefficients are fixed to $\Delta \theta _{\mathrm{LBW,}j} = 1$, corresponding to the incorporation of one gas-phase molecule into the surface.  The reverse transitions are assigned $\Delta \theta _{j, \mathrm{LBW}} = -1$.

The transitions between the two standing structures (SBW, SHB) usually incorporate desorption processes leading to $\Delta \theta_{\mathrm{SBW,SHB}} = \Delta \theta_{\mathrm{SHB,SBW}} = -1$.
The remaining free model parameters are obtained by fitting against the kMC simulation data. 

This is done by solving Equation \ref{eq:si_gn_model_ode} on a regular temperature and pressure grid (compare Figure \ref{fig:si_fig_acyclic}) with the initial condition of a fully covered LBW surface,
\begin{equation}
\label{eq:initial value}
     \vec{c}(0) = \left(\begin{array}{cc}
         1  \\
         0  \\
         0  \\
    \end{array} \right).
 \end{equation}

To be able to reliably solve the growth model differential equation \ref{eq:si_gn_model_ode} the interfacial length approximation $l_{ij} = c_i^{\alpha_i}c_j^{\alpha_j}$ needs regularization. In particular, each factor $c_i^{\alpha_i}$ was replaced by the smooth expression 
\begin{equation}
    c_i^{\alpha_i} \rightarrow c_i (c_i^2 + \epsilon)^{\frac{\alpha_i - 1}{2}}
\end{equation}
This regularization avoids the possible evaluation of small negative concentrations, which are nonphysical but because of numerical noise can occur during the solution process. The parameter $\epsilon$ is chosen to be small such that the regularized expression closely approximates $c_i^{\alpha_i}$ for physically relevant positive occupations $c_i$, while improving numerical stability near $c_i=0$.

The obtained model concentration trajectories are then used to compare against kMC data determined on the very same temperature and pressure grid with the same initial condition of a fully covered LBW surface. The residual for the least square fit at time $t_n$ is defined as
\begin{equation}
    \vec{R}_n = \vec{c}_\mathrm{model}(t_n, T, p ; f^\mathrm{nucl}, \Delta E^\mathrm{nucl}, \Delta \theta^\mathrm{nucl}, f^\mathrm{grow}, \Delta E^\mathrm{grow}, \Delta \theta^\mathrm{grow}, \alpha) - \vec{c}_\mathrm{kMC}(t_n, T, p),
\end{equation}

where $\vec{c}_\mathrm{model}$ is the solution to the initial value problem Equation \ref{eq:si_gn_model_ode} and \ref{eq:initial value}.
The optimal parameters are then determined by minimizing the least-squares objective
\begin{equation}
    \min \sum_n \left\lVert\vec{R}_n \right\rVert ^2.
\end{equation}

\section{Acyclicity penalty}

One specific feature of the introduced growth model is that the skew-symmetric growth term in Equation \ref{eq:si_gn_model_ode} introduces an effectively irreversible kinetic description. For each growth channel, the sign of the corresponding growth parameter defines a preferred direction of domain expansion, while an explicit competing reverse growth process is not included. Including the reverse processes would result in introducing additional fitting parameters making the fitting problem more complex with incremental improvement in fitting quality. Therefore, we stick to the irreversible model being aware of that detailed balance in each reaction channel $i \leftrightarrow j$ can not be fulfilled.

\begin{figure}
    \centering
    \includegraphics[width=10cm]{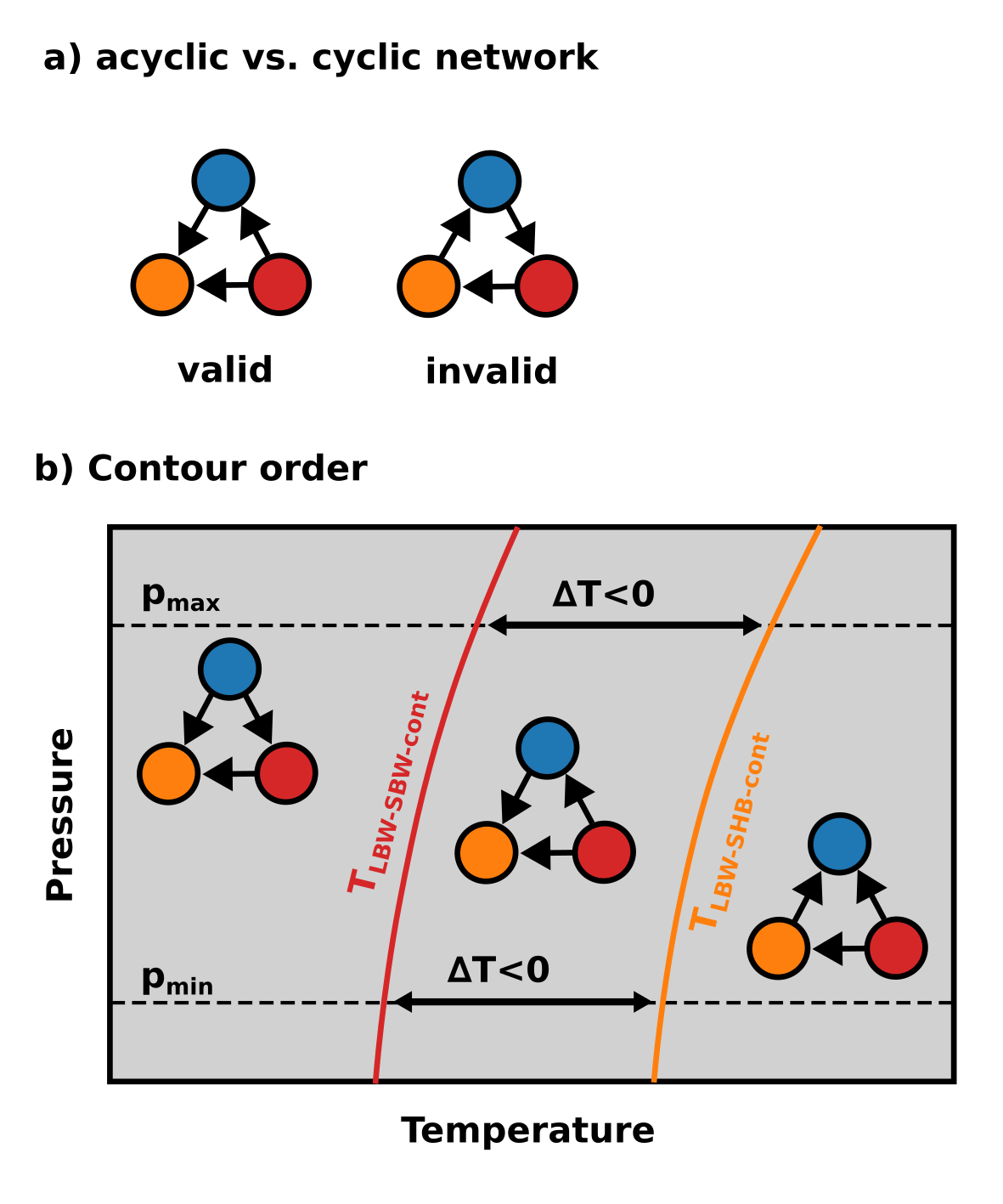}
    \caption{a) Comparison of a valid and invalid growth network. The right network exhibits a loop which can lead to oscillatory solutions. b) Description of the zero contour of the growth rates. Here the correct order is depicted where the contour $T_\mathrm{LBW\text{-}SBW\text{-}cont}$ (red line) is located at lower temperature compared to $T_\mathrm{LBW\text{-}SHB\text{-}cont}$ (orange line). The penalty to enforce the correct order is evaluated only at the minimum and maximum pressure. At these pressures the difference $\Delta T = T_\mathrm{LBW\text{-}SBW\text{-}cont} - T_\mathrm{LBW\text{-}SHB\text{-}cont}$ of the contour needs to be negative.}
    \label{fig:si_fig_acyclic}
\end{figure}

Because of this irreversible model it is possible to construct nonphysical cyclic reaction networks that would result in oscillatory solutions of Equation \ref{eq:si_gn_model_ode}. This nonphysical situation appears if the rate constants with positive sign in the growth matrix $k_{ij}$  result in a directed graph with a loop. In Figure \ref{fig:si_fig_acyclic}a exemplarily an invalid reaction network and a valid one is depicted.

To ensure that the fitting procedure does not produce nonphysical reaction networks it was necessary to include a simple penalty term in the least-square fitting procedure to circumvent loops in the growth rate matrix. 

In the case of the investigated 3-state model it is sufficient to ensure the correct order of the sign switch of the two rates $k_{\mathrm{LBW-SBW}}$ and $k_{\mathrm{LBW-SHB}}$ because the third rate $k_{\mathrm{SBW-SHB}}$ is positive at all relevant temperatures $T$ and pressures $p$. In particular, the system is acyclic at all relevant temperatures $T$ and pressures $p$ if the rate constant $k_\mathrm{LBW-SBW}$ switches sign before $k_\mathrm{LBW-SHB}$ while increasing temperature. This corresponds to ensure that the zero contour lines ($k_{\mathrm{LBW-SBW}}(T, p) = 0$ and $k_{\mathrm{LBW-SHB}}(T, p) = 0$) of the two rates have the correct order in the considered $T, p$ range.

The penalty term was designed to introduce a penalty in the least-square residual if the order is violated at the pressure boundaries $p_{\mathrm{min}/\mathrm{max}}$.

For that the zero contour line as a function of pressure is needed (Figure \ref{fig:si_fig_acyclic}) and can be calculated setting Equation \ref{eq:growth_rate_const} to zero and rearrange terms, which leads to
\begin{equation}
    T_{ij\text{-}\mathrm{cont}}(p) =  \frac{\Delta E^\mathrm{grow}_{ij} - \Delta E^\mathrm{grow}_{ji}}{k_\mathrm{B} \ln\left(\frac{f_{ij}^\mathrm{grow}}{f_{ji}^\mathrm{grow}} p^{\Delta \theta_{ij} - \Delta \theta_{ji}}\right)}.
\end{equation}
The penalty terms included to the least-square residual are then given by a softplus function of the difference of the contour lines which returns a large positive value if the order is wrong and if it is correct the penalty is vanishing. It is given by
\begin{equation}
    \lambda(p_i) = \ln\left(1 + e^{T_{\mathrm{LBW\text{-}SBW\text{-}cont}}(p_i) -T_\mathrm{LBW\text{-}SHB\text{-}cont}(p_i)}\right)
\end{equation}
where $p_i$ are the minimum and maximum pressure in the system ($p_\mathrm{min} = 10^{-8} \mathrm{~bar}$, $p_\mathrm{max} = 10^{-6} \mathrm{~bar}$). The adapted least-square objective is the given by

\begin{equation}
    \min \left[\sum_n \left\lVert\vec{R}_n \right\rVert ^2 + \lambda(p_\mathrm{min})^2 + \lambda(p_\mathrm{max})^2 \right].
\end{equation}

\section{Obtained fitting parameters}

We use the \textit{scipy }Python package \cite{virtanen2020SciPy10Fundamental} for performing the least-squares optimizations.
The obtained parameters are given by

\begin{equation}
    f^\mathrm{nucl} = \left(\begin{array}{ccc}
         \cdot& 5.075 \times10^{7} & 4.852 \times10^{8}\\
        2.762 \times10^{-6} & \cdot&2.235 \times10^{-43}\\
         4.665 \times10^{-1}&3.446 \times10^{5} & \cdot
    \end{array} \right) \mathrm{~s^{-1}}
\end{equation}

\begin{equation}
    \Delta E^\mathrm{nucl} =\left(\begin{array}{ccc}
        \cdot & 0.386 & 0.526  \\
         0.543& \cdot&  4.768\\
         2.379& 1.347 & \cdot \\
    \end{array} \right) \mathrm{~eV}
\end{equation}

\begin{equation}
    \Delta \theta^\mathrm{nucl} =\left(\begin{array}{ccc}
        \cdot &1 & 1  \\
         -1 & \cdot&  -1\\
         -1 & -1 & \cdot \\
    \end{array} \right)
\end{equation}
\begin{equation}
    \left(\begin{array}{c}
        \alpha_{\mathrm{LBW}}\\
         \alpha_{\mathrm{SBW}}\\
         \alpha_{\mathrm{SHB}}
    \end{array}\right) = \left(\begin{array}{c}
        1.384\\
         1.020\\
         1.005
    \end{array}\right)
\end{equation}

\begin{equation}
    f^\mathrm{grow} = \left(\begin{array}{ccc}
         \cdot& 1.829 \times10^{8} & 3.271 \times10^{9}\\
        7.975 \times10^{38} & \cdot&1.208 \times10^{40}\\
         6.893 \times10^{40}&2.293 \times10^{24} & \cdot
    \end{array} \right)
\end{equation}

\begin{equation}
    \Delta E^\mathrm{grow} =\left(\begin{array}{ccc}
        \cdot & 0.272 & 0.363  \\
         4.295& \cdot&  4.301\\
         4.471& 5.150 & \cdot \\
    \end{array} \right)
\end{equation}

\begin{equation}
\label{eq:theta_values}
    \Delta \theta^\mathrm{grow} =\left(\begin{array}{ccc}
        \cdot &1 & 1  \\
         -1 & \cdot&  -1\\
         -1 & -1 & \cdot \\
    \end{array} \right)
\end{equation}

\begin{figure}[ht]
    \centering
    \includegraphics[width=\textwidth]{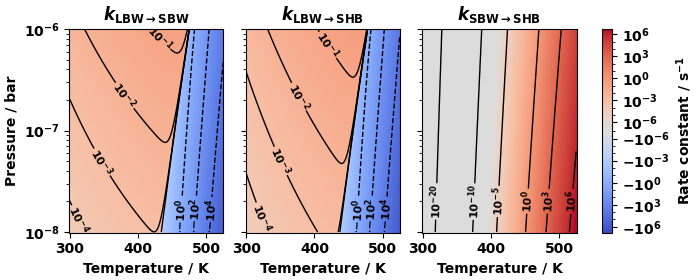}
    \caption{Temperature and pressure dependence of the three different growth rates.}
    \label{fig:si_fig_growth_rates}
\end{figure}

\begin{figure}[ht]
    \centering
    \includegraphics[width=\textwidth]{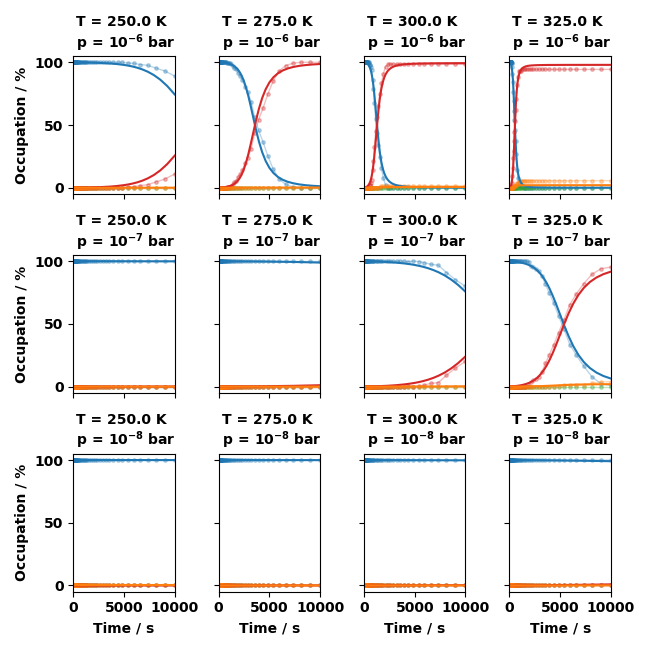}
    \caption{Obtained model occupation trajectories of LBW (blue), SBW (red) and SHB (orange) fitted on the corresponding kMC data (circle markers).}
    \label{fig:fig250_fit}
\end{figure}

\begin{figure}[ht]
    \centering
    \includegraphics[width=\textwidth]{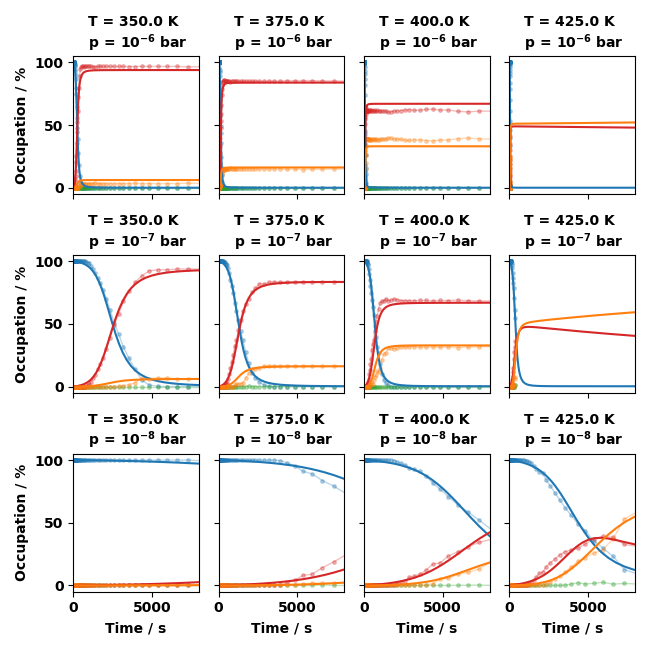}
    \caption{Obtained model occupation trajectories of LBW (blue), SBW (red) and SHB (orange) fitted on the corresponding kMC data (circle markers).}
    \label{fig:fig350_fit}
\end{figure}

\begin{figure}[ht]
    \centering
    \includegraphics[width=\textwidth]{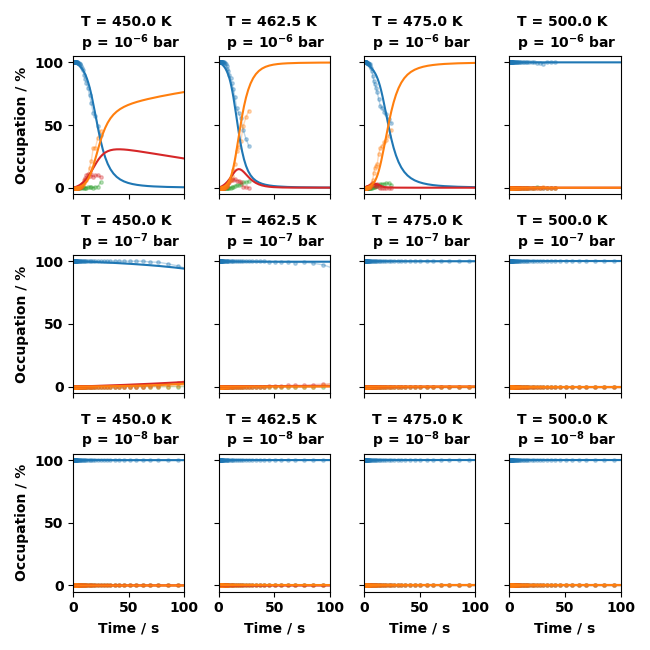}
    \caption{Obtained model occupation trajectories of LBW (blue), SBW (red) and SHB (orange) fitted on the corresponding kMC data (circle markers).}
    \label{fig:fig450_fit}
\end{figure}

\begin{figure}[ht]
    \centering
    \includegraphics[width=\textwidth]{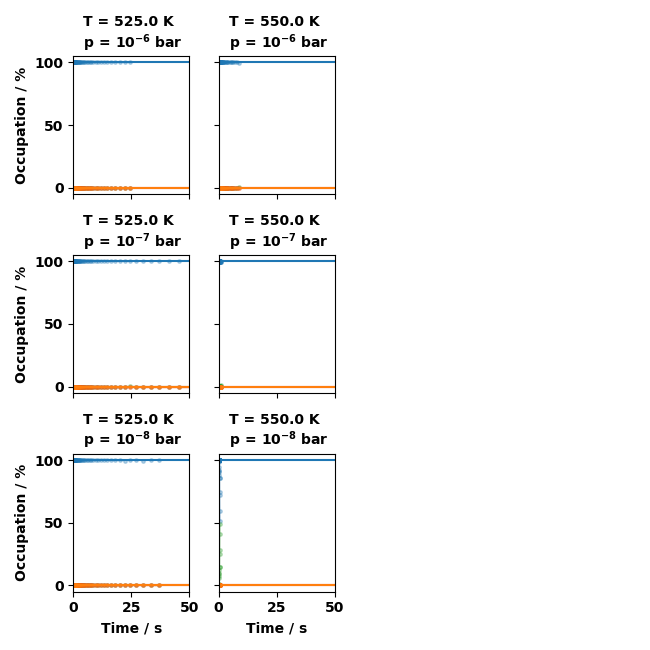}
    \caption{Obtained model occupation trajectories of LBW (blue), SBW (red) and SHB (orange) fitted on the corresponding kMC data (circle markers).}
    \label{fig:fig525_fit}
\end{figure}

The temperature and pressure dependence of the growth rate constants is shown in Figure \ref{fig:si_fig_growth_rates}. Figures \ref{fig:fig250_fit}-\ref{fig:fig525_fit} show the occupation trajectories obtained from the parametrized effective growth model together with the corresponding kMC simulation data. The comparison demonstrates that the parameterized model captures the main features of the kMC trajectories over most of the investigated parameter range.

\section{Solving the optimal control problem}

To solve the optimal control problem the \textit{yapss} Python package was used. This software uses a pseudo-spectral approach \cite{patterson2014GPOPSIIMATLABSoftwarea} which expresses the continuous optimal control problem in Lagrange polynomials transforming the problem into a finite-dimensional nonlinear programming problem. This optimization problem is then solved by the \textit{Ipopt} \cite{wachter2006ImplementationInteriorpointFilter} package.

To efficiently incorporate all the mentioned constraints, the state vector $\vec{s}$ of the system is extended by the temperature $T$ and logarithmic pressure $q = \log_{10}p$ and its derivatives and is given by

\begin{equation}
    \vec{s} = \left(\begin{array}{c}
        c_\mathrm{LBW}   \\
        c_\mathrm{SBW}   \\
         c_\mathrm{SHB}  \\
         T  \\
         q  \\
         \dot{T}  \\
         \dot{q}  \\
    \end{array} \right)
\end{equation}
The control variable $\vec{u}$ are the second derivatives of the $T$ and $q$.

\begin{equation}
    \vec{u}(t) = \left(\begin{array}{c}
           \ddot{T}(t)\\
            \ddot{q}(t)\\
    \end{array} \right)
\end{equation}

The corresponding differential equation is now

\begin{equation}
\label{eq:extended_ode_eq}
    \frac{\diff }{\diff t} \vec{s} = \left(\begin{array}{c}
          \frac{\diff }{\diff t} c_\mathrm{LBW}\\
         \frac{\diff }{\diff t} c_\mathrm{SBW}\\
        \frac{\diff}{\diff t}  c_\mathrm{SHB} \\
         \dot{T} \\
         \dot{q}\\
         \ddot{T}\\
         \ddot{q} \\
    \end{array}\right) \quad \text{with} \quad \vec{s}(0) = \left(\begin{array}{l}
          c_\mathrm{LBW} = 1\\
          c_\mathrm{SBW} = 0 \\
         c_\mathrm{SHB} = 0\\
         T_\mathrm{start} = 525~\mathrm{K}\\
         q_\mathrm{start} = \log_{10}(10^{-7} \mathrm{~bar})\\
        \dot{T}_\mathrm{start} = 0 ~\mathrm{Ks^{-1}}\\
         \dot{q}_\mathrm{start} = 0~ \log_{10}(\mathrm{bar})\mathrm{s}^{-1}\\
    \end{array} \right)
\end{equation}
where the derivatives of the concentrations are given by \ref{eq:si_gn_model_ode}. In the initial condition we demand vanishing first derivatives in temperature and pressure at $t=0$.

Additionally to the initial conditions given in \ref{eq:extended_ode_eq} we also enforce vanishing derivatives at final time, $\dot{T}(t_\mathrm{f}) = 0$ and $\dot{q}(t_\mathrm{f})=0$. 

To keep the solutions of the optimal control problem within reasonable bounds the temperature and logarithmic pressure and its derivatives are bound to minimum and maximum values. All used values are listed in Table \ref{tab:bounds}.

\begin{table}[]
    \centering
\caption{Bounds for the state and control variables of the optimal control problem}
\label{tab:bounds}
    \begin{tabular}{|c|l|l|}
    \hline
          Variable&Min&  Max\\
          \hline
          $T$&300 K&  525 K\\
 $q$&  $\log_{10}(10^{-8} \mathrm{~bar})$&$\log_{10}(10^{-6} \mathrm{~bar})$\\
 $\dot{T}$& $-1 \mathrm{~Ks^{-1}}$&$1 \mathrm{~Ks^{-1}}$\\
 $\dot{q}$& $-0.01 \log_{10}(\mathrm{bar})s^{-1}$&$0.01 \log_{10}(\mathrm{bar})s^{-1}$\\
 $\ddot{T}$& $-0.2 \mathrm{~Ks^{-2}}$&$0.2 \mathrm{~Ks^{-2}}$\\
 $\ddot{q}$& $-0.002 \log_{10}(\mathrm{bar})s^{-2}$&$0.002 \log_{10}(\mathrm{bar})s^{-2}$\\
 \hline
 \end{tabular}
           
\end{table}

The objective functional of the optimal control problem is given in Equation \ref{eq:oct_objective} of the main text as

\begin{equation*}
    J[T, q] = -c_\mathrm{SBW}(t_\mathrm{f}) + \int_0^{t_\mathrm{f}} \left[ \alpha_T \ddot{T}^2(t) + \alpha_q \ddot{q} ^2 (t) \right] \diff t.
\end{equation*}
The integral term acts as a regularization term that promotes smooth temperature and pressure protocols. The regularization strengths were set to $\alpha_T = 0.25$ and $\alpha_q= 2500$, which was found to yield smooth and physically reasonable optimized trajectories for both temperature and pressure.  
The substantially larger value of $\alpha_q$ compensates for the different numerical scales of the two control variables. Without this rescaling, the regularization would predominantly affect the control variable with the larger numerical magnitude, while variations in the smaller-scale variable would be penalized much less strongly.
\FloatBarrier
\newpage
\section*{References}
\printbibliography[heading=none]
\end{refsection}